\documentclass{IEEEtran}
\usepackage{amsmath,amsfonts}
\usepackage{algorithm}
\usepackage{algpseudocode}
\usepackage{array}
\usepackage[caption=false,font=normalsize,labelfont=sf,textfont=sf]{subfig}
\usepackage{textcomp}
\usepackage{stfloats}
\usepackage{url}
\usepackage{verbatim}
\usepackage{graphicx}
\usepackage{cite}
\usepackage{color}
\usepackage{xcolor}
\usepackage{multirow}
\usepackage{mathabx}
\allowdisplaybreaks

\begin{document}
\title{Privacy-Preserving Utilization of Distribution System Flexibility for Enhanced TSO-DSO Interoperability: A Novel Machine Learning-Based Optimal Power Flow Approach}

\author{Burak~Dindar,~\IEEEmembership{Graduate Student Member,~IEEE,}
          ~Can~Berk~Saner,~\IEEEmembership{Member,~IEEE,}
          ~Hüseyin~K.~Çakmak,
          and~Veit~Hagenmeyer,~\IEEEmembership{Member,~IEEE}
          
          \thanks{This work was partly conducted within the framework of the Helmholtz Program Energy System Design (ESD) and the DigIPlat project, which received funding in the framework of the joint programming initiative ERA-Net Smart Energy Systems’ focus initiative Digital Transformation for the Energy Transition, with support from the European Union’s Horizon 2020 research and innovation program under grant agreement No 883973.
          
          Burak Dindar, Hüseyin K. Çakmak and Veit Hagenmeyer are with the Institute for Automation and Applied Informatics, Karlsruhe Institute of Technology, 76131 Karlsruhe, Germany, (e-mail:burak.dindar@kit.edu; hueseyin.cakmak@kit.edu; veit.hagenmeyer@kit.edu).
          
          Can Berk Saner is with the Department of Electrical and Computer Engineering, National University of Singapore, Singapore 117581, (e-mail: sanerc@u.nus.edu).
          
}}

\maketitle

\begin{abstract}
Due to the transformation of the power system, the effective use of flexibility from the distribution system (DS) is becoming crucial for efficient network management. Leveraging this flexibility requires interoperability among stakeholders, including Transmission System Operators (TSOs) and Distribution System Operators (DSOs). However, data privacy concerns among stakeholders present significant challenges for utilizing this flexibility effectively. To address these challenges, we propose a machine learning (ML)-based method in which the technical constraints of the DSs are represented by ML models trained exclusively on non-sensitive data. Using these models, the TSO can solve the optimal power flow (OPF) problem and directly determine the dispatch of flexibility-providing units (FPUs)—in our case, distributed generators (DGs)-in a single round of communication. To achieve this, we introduce a novel neural network (NN) architecture specifically designed to efficiently represent the feasible region of the DSs, ensuring computational effectiveness. Furthermore, we incorporate various PQ charts rather than idealized ones, demonstrating that the proposed method is adaptable to a wide range of FPU characteristics. To assess the effectiveness of the proposed method, we benchmark it against the standard AC-OPF on multiple DSs with meshed connections and multiple points of common coupling (PCCs) with varying voltage magnitudes. The numerical results indicate that the proposed method achieves performant results while prioritizing data privacy. Additionally, since this method directly determines the dispatch of FPUs, it eliminates the need for an additional disaggregation step. By representing the DSs technical constraints through ML models trained exclusively on non-sensitive data, the transfer of sensitive information between stakeholders is prevented. Consequently, even if reverse engineering is applied to these ML models, no sensitive data can be extracted. This allows for the utilization of DS flexibility in network management without compromising data privacy, thereby enhancing interoperability among stakeholders.
\end{abstract}


\section{Introduction}

\IEEEPARstart{W}{ith} the rapid transformation of the power system, the number of  flexibility-providing units (FPUs), such as distributed generators (DGs) connected to distribution system (DS) is steadily increasing. The inherent fluctuations associated with DGs complicate the management not only of the DS but also of the transmission system (TS) \cite{jia2014hierarchical}. On the other hand, the flexibility provided by DSs can be effectively leveraged for the provision of ancillary services, contributing to the stability and reliability of the entire power system \cite{ringelstein2022methodology}. As the number of FPUs continues to rise, the necessity for effectively managing these flexibilities is becoming increasingly critical. However, the effective utilization of these flexibilities necessitates a high level of coordination between Transmission System Operators (TSOs) and Distribution System Operators (DSOs) \cite{gerard2016basic}.

In recent years, increasing the coordination between TSOs and DSOs and the utilization of DSs flexibility in ancillary services have garnered significant attention from researchers, leading to numerous studies focused on developing innovative coordination schemes \cite{givisiez2020review, dai2024real, lind2023evaluation}. These schemes typically require specific data exchanges between TSOs and DSOs in predefined formats. However, despite existing agreements governing such data transfers, the implementation of these coordination schemes in real-world projects faces numerous challenges and barriers \cite{migliavacca2019tso}. One major issue is the unwillingness of stakeholders, such as TSOs and DSOs, to share essential data \cite{ziesemann2023challenges}. Current bilateral agreements often fail to address key concerns, such as data leakage, which can lead to the unintended disclosure of sensitive information. For instance, coordination schemes may expose DS system topology data (e.g., line parameters) or customer-specific load data, jeopardizing both commercially sensitive information and the privacy of individual customers. These concerns hinder interoperability and pose a significant challenge to the efficient operation of the power system \cite{habibi2022privacy}. Therefore, the primary objective of the present paper is to eliminate the exchange of commercially and personally sensitive data between TSOs and DSOs while ensuring overall data protection and privacy.

In this context, differential privacy (DP) has been investigated as a method for protecting sensitive data in power systems \cite{mak2019privacy}. For example, DP has been applied to obscure transmission line and transformer parameters during data exchange in power grids \cite{fioretto2019differential}. Similarly, customer load data in distributed OPF has been protected using DP techniques \cite{mak2020privacy}. In this method, noise is added to the data to prevent the exposure of sensitive information. While this approach enhances data protection, the introduction of noise can pose significant challenges in complex optimization algorithms such as OPF, potentially leading to infeasible solutions \cite{fioretto2018constrained}. This, in turn, limits the effective utilization of FPU potential. As highlighted in \cite{fioretto2019differential}, additional mechanisms are necessary to maintain high accuracy while preserving data privacy. However, implementing such mechanisms introduces extra computational overhead. Moreover, existing studies primarily focus on protecting specific types of data, without offering comprehensive solutions to safeguard all sensitive data simultaneously.

Another commonly used approach to ensure comprehensive data privacy in TSO-DSO interactions is the distributed OPF method \cite{dai2025advancing}. In this approach, the OPF problem is decomposed into sub-problems to prevent the need for sharing complete grid models. However, it still requires the exchange of sensitive information such as complex voltages and/or active and reactive power flows at tie-lines between neighboring regions. While this method allows for the effective integration of FPUs from DSs into the TSO’s OPF, it suffers from several limitations \cite{jiang2022risk}. Firstly, the approach relies on iterative information exchanges between regions to achieve convergence, which significantly increases communication complexity. Secondly, as the number of DSOs in the system grows, the number of iterations and the time required for convergence rise considerably, posing scalability challenges \cite{dai2024ensuring}. Additionally, these methods often model FPUs using idealized rectangular PQ characteristics, failing to capture the diversity of real-world PQ capabilities.

Additionally, wide range of approaches focuses on the concept of PQ capability charts to ensure data privacy in TSO-DSO coordination \cite{silva2018estimating, capitanescu2018tso}. In this approach, the DSO calculates the aggregated flexibility at the TSO-DSO interface within the PQ domain \cite{churkin2023tracing, wang2021aggregate}. This PQ region, often represented as a polygon, defines the feasible operating region (FOR) of the DS \cite{contreras2019time}. The TSO can then leverage these aggregated flexibilities for power system operations without the need to exchange sensitive data, such as the grid model \cite{vijay2022complex, fortenbacher2020reduced}. 

In addition to the advantages related to data privacy, the PQ capability chart approach has a key limitation \cite{usman2023novel}: Specifically, the cost associated with any point on the PQ chart reflects the aggregate costs of various DGs, making it difficult to directly incorporate the cost implications of a TSO's selected point in the analysis \cite{sarstedt2022monetarization}. Consequently, an additional disaggregation problem must be addressed to account for the individual costs of DGs  \cite{chen2021leveraging}. For instance, in \cite{polymeneas2016aggregate}, a two-level hierarchical optimization scheme is proposed, where DGs are first aggregated, a multi-step optimal power flow (OPF) is performed, and then an optimization-based disaggregation problem is solved. Similarly, in \cite{fruh2022coordinated}, a top-down disaggregation process across voltage levels, based on a linear OPF model, is introduced and tested on a real distribution system. As illustrated, the PQ capability chart approach necessitates solving the disaggregation problem to effectively utilize aggregated flexibility in ancillary services, which introduces additional workload and requires iterative communication between TSOs and DSOs.

Another important aspect to consider regarding the PQ capability chart approach is the simplifications often employed in the method. In many studies, it is assumed that the DS is connected to the TS through a single point of common coupling (PCC), and radial test systems are utilized \cite{kalantar2019characterizing, riaz2021modelling, tan2020estimating}. However, in reality, many DSs are operated in a meshed configuration, with multiple PCCs between TSOs and DSOs. Germany is a prominent example of this complexity; its 110 kV grid is meshed, connected to the TS via multiple PCCs, and managed by DSOs \cite{stock2018optimal}. Moreover, a common assumption in the literature is that the voltage at the TSO-DSO interface, i.e., the PCC, remains constant \cite{xu2022voltage}. However, in practical scenarios, the voltage at the PCC fluctuates depending on dispatch decisions. Assuming a constant voltage at the PCC can lead to an inaccurate assessment of DS flexibility potential, ultimately limiting its effective utilization. These simplifications and assumptions hinder the practical application of the PQ capability chart approach in real-world scenarios.

Considering the aforementioned challenges, our previous works \cite{dindar2023tso, dindar2024machine} introduced a machine learning (ML)-based methodology to integrate DGs located within the DS into the OPF problem, which is solved by the TSO, while maintaining data privacy. Although various coordination schemes exist, ENTSO-E asserts that TSOs hold the primary responsibility for overall system security, while DSOs are tasked with ensuring the secure operation of their respective DSs \cite{entso2015towards}. In alignment with these responsibilities, our approach involves the DSO developing ML models that encapsulate the technical constraints of the DS based solely on the active and reactive power outputs of the DGs and the voltage magnitude at the PCC. By training ML models with this limited dataset, which comprising only information already known and shared between the TSO and DSO, commercially (e.g., system topology) and individual (e.g., customer load profiles) sensitive data is inherently protected, as these details are excluded from the dataset used for training. It is important to note that while ML models are generally susceptible to model inversion attacks, the proposed method ensures that even if reverse engineering is applied, no sensitive information is exposed, as the ML models are trained exclusively with non-sensitive data. Once trained, these ML models are transferred to the TSO, which subsequently utilizes them to solve the OPF problem, including the direct determination of DG dispatch within a single communication round. This approach not only guarantees data privacy—by enabling the DSO to share only ML models trained on non-sensitive data—but also ensures that the overall system is managed by the TSO in compliance with ENTSO-E's operational framework. Simultaneously, the method considers the technical constraints of both the DS (through ML models) and the TS, facilitating a secure and coordinated operation.

In the present paper, we significantly enhance our previously proposed method by addressing the aforementioned challenges: The new approach extends the application of ML-based privacy-preserving OPF to not only a single DS but also to multiple DSs, even when there are multiple PCCs involved. Furthermore, we introduce a novel tailored neural network (NN) to accurately and efficiently represent the feasible operating region of the DSs. To generate the necessary data for creating the ML models, the Latin hypercube sampling (LHS) method is employed. Additionally, to demonstrate the adaptability of the proposed method in handling diverse FPUs with varying PQ characteristics, we do not limit our study to DGs modeled with simple rectangular PQ charts. Instead, we also consider PQ charts with different characteristics. The LHS-based dataset generation is accordingly adjusted to reflect these varied characteristics. Finally, with the proposed method, instead of defining a PQ chart at the TSO-DSO interface, the flexibility of the DSs can be directly utilized by the TSO in power system management.

The key contributions of the present paper are as follows:

\begin{itemize}
\item Direct integration of FPUs into the OPF problem, enabling DG dispatch within a single communication round, thus eliminating the need for an additional disaggregation.
\item Effective incorporation of DS flexibility in complex meshed systems, accommodating scenarios with multiple DSs and multiple PCCs, including treating PCC voltage as a variable to enhance DS flexibility.
\item Broad adaptability, allowing integration of FPUs with diverse PQ characteristics.
\item The novel NN architecture enables efficient and accurate representation of the DSs' feasible operating region, improving computational performance.
\item Overall, the proposed method ensures the effective utilization of flexibility from DSs for network management, while maintaining data privacy and respecting the operational limits of both TSs and DSs.
\end{itemize}

The rest of the paper is organized as follows: In Section \ref{sec:method}, we present the proposed methodology. In Section \ref{sec:dataset_creation}, we introduce the dataset creation technique. Subsequently, in Section \ref{sec:ml_models}, we detail the representation of the DSs with ML models. Then, we benchmark the proposed method against the standard AC-OPF to evaluate its effectiveness various different case studies in Section \ref{sec:case}. Finally, we present our conclusions in Section \ref{sec:conclusion}.

\section{Overview of the Proposed Methodology}\label{sec:method}

To set the notation in this paper, parameters are denoted by standard letters ($a, A$), and variables are represented using boldface letters ($\boldsymbol{a}$, $\boldsymbol{A}$), while sets are represented by calligraphic letters ($\mathcal{A}$). Matrices are denoted by uppercase ($A$), while scalar and (column) vector variables/parameters are presented in lowercase letters ($a$). Furthermore, functions are expressed by $A(\cdot)$. The $n$-th element of a vector $a$ is denoted as $a^{(n)}$, and  the $n$-th row of a matrix $A$ is denoted as $A^{(n,:)}$. Moreover, the element at position $(i, j)$ in a matrix is expressed by $A^{(i, j)}$. Finally, the symbols $\preceq$ and $\succeq$ are used for element-wise $\leq$ and $\geq$ comparisons, respectively.

\subsection{Formulation of the Standard AC-OPF}\label{sec:std_opf}

In the present paper, we consider an integrated power system with a total of $n_{\mathrm{b}}$ buses,  comprising a transmission system (TS) with $n_{\mathrm{g}}$ conventional generator, and $n_{\mathrm{b,ts}}$ buses, as well as $n_{\mathrm{ds}}$ distribution systems (DSs), where the $j$-th DS contains $n_{\mathrm{dg},j}$ distributed generators (DGs). Note that some DSs have multiple points of common coupling (PCCs) with the TS. Following this consideration we can define the standard AC-OPF as follows:

\begin{subequations}\label{eq:opf_dsotso}
\begin{align}
\min_{\substack{\boldsymbol{\widehat{v}},\boldsymbol{\widehat{\theta}}, \\ \boldsymbol{\widecheck{p}}_{\mathrm{g}}, \boldsymbol{\widecheck{q}}_{\mathrm{g}}, \\ \boldsymbol{p}_{\mathrm{dg},j}, \\ \boldsymbol{q}_{\mathrm{dg},j}}} \quad & \sum_{i = 1}^{n_{\mathrm{g}}}C_{i}(\boldsymbol{{\widecheck{p}}}_{\mathrm{g}}^{(i)}) + \sum_{j = 1}^{n_{\mathrm{ds}}}\sum_{k = 1}^{n_{\mathrm{dg,}j}}C_{jk}(\boldsymbol{p}_{\mathrm{dg,}j}^{(k)}) \label{eq:opf_dsotso0}\\
\text{s.t.} \quad & {G}_{\mathrm{P}}(\boldsymbol{\widehat{v}},  \boldsymbol{\widehat{\theta}}; \widehat{Y}) + \widehat{p}_{\mathrm{d}} - K \boldsymbol{\widecheck{p}}_{\mathrm{g}} - \sum_{j=1}^{n_{\mathrm{ds}}} H_{j} \boldsymbol{p}_{\mathrm{dg,}j} = 0, \label{eq:opf_dsotso1} \\
& {G}_{\mathrm{Q}}(\boldsymbol{\widehat{v}},  \boldsymbol{\widehat{\theta}}; \widehat{Y}) + \widehat{q}_{\mathrm{d}} - K \boldsymbol{\widecheck{q}}_{\mathrm{g}} - \sum_{j=1}^{n_{\mathrm{ds}}} H_{j} \boldsymbol{q}_{\mathrm{dg,}j} = 0, \label{eq:opf_dsotso2} \\
& G_{\mathrm{line}}(\boldsymbol{\widehat{v}},  \boldsymbol{\widehat{\theta}}; \widehat{Y}) \preceq \widehat{l}_{\mathrm{line, max}}, \label{eq:opf_dsotso3} \\
& \widehat{v}_{\mathrm{min}} \preceq \boldsymbol{\widehat{v}} \preceq \widehat{v}_{\mathrm{max}}, \ \widehat{\theta}_{\mathrm{min}} \preceq \boldsymbol{\widehat{\theta}} \preceq \widehat{\theta}_{\mathrm{max}}, \label{eq:opf_dsotso4} \\
& \widecheck{p}_{\mathrm{g, min}} \preceq \boldsymbol{\widecheck{p}}_{\mathrm{g}} \preceq \widecheck{p}_{\mathrm{g, max}}, \ \widecheck{q}_{\mathrm{g, min}} \preceq \boldsymbol{\widecheck{q}}_{\mathrm{g}} \preceq \widecheck{q}_{\mathrm{g, max}}, \label{eq:opf_dsotso5} \\
& p_{\mathrm{dg},j, \mathrm{min}} \preceq \boldsymbol{p}_{\mathrm{dg}, j} \preceq p_{\mathrm{dg},j, \mathrm{max}}, \ \forall j \in \{1,., n_{\mathrm{ds}}\}, \label{eq:opf_dsotso6} \\
& q_{\mathrm{dg},j, \mathrm{min}} \preceq \boldsymbol{q}_{\mathrm{dg}, j} \preceq q_{\mathrm{dg},j, \mathrm{max}}, \ \forall j \in \{1,.., n_{\mathrm{ds}}\}. \label{eq:opf_dsotso7}
\end{align}
\end{subequations}

For clarity and ease of reference, we adopt the following notation: variables associated with the integrated system (including both TS and DS) are denoted with a hat ($\boldsymbol{\widehat{a}}$), variables associated solely with the TS are denoted with an inverted hat ($\boldsymbol{\widecheck{a}}$), and variables related exclusively to the DS are presented without a hat ($\boldsymbol{a}$). For example, $\boldsymbol{\widehat{v}}$, represents the voltage magnitudes of all buses in the integrated system, while $\boldsymbol{\widecheck{v}}$ refers only to the TS buses.

Following this convention, $\boldsymbol{\widehat{v}}, \boldsymbol{\widehat{\theta}}, \widehat{p}_{\mathrm{d}},$ and $\widehat{q}_{\mathrm{d}} \in \mathbb{R}^{n_{\mathrm{b}}}$ represent the vectors of bus voltage magnitude, voltage angle, active and reactive power demand vectors respectively, for the integrated system, which include both TS and DSs. $\widehat{Y} \in \mathbb{R}^{{n}_{\mathrm{b}} \times {n}_{\mathrm{b}}}$ denotes the bus admittance matrix. $\boldsymbol{\widecheck{p}}_{\mathrm{g}}, \boldsymbol{\widecheck{q}}_{\mathrm{g}} \in \mathbb{R}^{n_{\mathrm{b,ts}}}$ are the vectors of active and reactive power generation for the TS buses. $K$ is the $n_{\mathrm{b}} \times n_{\mathrm{b,ts}}$ \textit{transmission generation connection matrix} such that the element $(t,v)$ is one if this element is located inside the TS, and zero otherwise. The vectors $\boldsymbol{p}_{\mathrm{dg},j}, \boldsymbol{q}_{\mathrm{dg},j} \in \mathbb{R}^{n_\mathrm{dg},j}$ correspond to the active and reactive power generation of the DGs in the $j$-th DS. $H_{j}$ is the $n_{\mathrm{b}} \times n_{\mathrm{dg},j}$ \textit{distributed generation connection matrix} such that the element $(m,n)$ is one if $n$-th DG of the $j$-th DS is located at bus $m$, and zero otherwise. It is important to note that the size of the vectors $\boldsymbol{\widecheck{p}}_{\mathrm{g}}$ and $\boldsymbol{\widecheck{q}}_{\mathrm{g}}$ corresponds to the number of TS buses, $n_{\mathrm{b,ts}}$, while the size of the vectors $\boldsymbol{p}_{\mathrm{dg},j}$ and $\boldsymbol{q}_{\mathrm{dg},j}$ corresponds to the number of DGs, $n_{\mathrm{dg},j}$.

Moreover, in \eqref{eq:opf_dsotso0}, the objective function minimizes the total cost of generation dispatch, including DGs. Here, $C_{i}(\cdot)$ represents the cost of active power generation at bus $i$, similarly, $C_{jk}(\cdot)$ represents the cost of active power generation for the $k$-th DG in the $j$-th DS. Without loss of generality, we consider a standard quadratic cost function for both functions, expressed as $C_{l}(\boldsymbol{p}) = a_{l}\boldsymbol{p}^{2}+b_{l}\boldsymbol{p}+c_{l}$. Note that, in this integrated system, we assume that the first $n_{\mathrm{g}}$ buses are associated with conventional generators for notational convenience. Equations \eqref{eq:opf_dsotso1} and \eqref{eq:opf_dsotso2} represent the active and reactive balance equations, where $G_{P}(\cdot)$ and $G_{Q}(\cdot)$ are the corresponding functions. In \eqref{eq:opf_dsotso3}, $G_{line}(\cdot)$ denotes the line apparent power flows, which is bounded by the line flow limit vector $\widehat{l}_{\mathrm{line, max}}$. Finally, \eqref{eq:opf_dsotso4} - \eqref{eq:opf_dsotso7} establish the upper and lower bounds for the respective variables.

Examining Equation \eqref{eq:opf_dsotso}, it becomes evident that the utilization of flexibility from DSs in network management requires access to sensitive data for the entire system. For instance, the admittance matrix $\widehat{Y}$ encapsulates the topology of the system, while the demand vectors $\widehat{p}_{\mathrm{d}}$ and $\widehat{q}_{\mathrm{d}}$ contain load data. Typically, since the OPF problem is solved by the TSO, DSOs are reluctant to share such sensitive data with TSOs. To address this issue, in the present paper, we introduce a novel AC-OPF formulation designed to eliminate the need for sensitive data exchange between TSOs and DSOs. This new formulation allows for the effective use of DS flexibility in power system management while maintaining data privacy.

\subsection{Formulation of the ML-Based Privacy-Preserving AC-OPF}\label{sec:pro_opf}

In our novel AC-OPF formulation, the primary goal is to prevent the exchange of sensitive data between TSOs and DSOs. To achieve this, we separate the DS-related variables and parameters from the integrated system. As previously described, we assume that there are $n_{\mathrm{ds}}$ distribution systems, and the $j$-th DS contains $n_{\mathrm{dg},j}$ distributed generators, where $j \in \{1, 2, \dots, n_{\mathrm{ds}}\}$. We extend this setup by assuming that each distribution system $j$ is connected to specific buses $\{s_{j, 1}, s_{j, 2}, \dots, s_{j, r_j}\}$ (i.e., the points of common coupling (PCCs)) of the TS, where $r_j$ denotes the number of PCCs for the $j$-th DS. These PCCs in the TS are treated as \textit{empty} buses, meaning these buses do not have any directly connected generators or loads.

Accordingly, we model each DS at the corresponding PCCs as dependent active and reactive power injections. These injections represent the power flows at the PCCs. For instance, a DS with a single PCC is modeled at that PCC, while a DS with multiple PCCs is represented by separate active and reactive power flows at each respective PCC. This representation depends on the vector $\boldsymbol{v}_{j} \in \mathbb{R}^{r_{j}}$, which consists of the voltage magnitudes at the PCCs (${s_{j,1}, s_{j,2}, \dots, s_{j,r_j}}$) of the $j$-th DS. It also depends on active and reactive power generation vectors of DGs, $\boldsymbol{p}_{\mathrm{dg},j}$ and $\boldsymbol{q}_{\mathrm{dg},j}$ for the $j$-th DS. For convenience, we concatenate these variables into a single vector $\boldsymbol{x}_{j} = {\begin{bmatrix}
 \boldsymbol{v}_{j}^{\top}&{\boldsymbol{p}_{\mathrm{dg},j}}^{\top}  &{\boldsymbol{q}_{\mathrm{dg},j}}^{\top}\end{bmatrix}}^{\top} \in \mathbb{R}^{n_{j}}$, where $n_{\mathrm{j}} = r_j+ 2n_{\mathrm{dg},j}$. With this setup, we can define the proposed privacy-preserving AC-OPF as follows:

\begin{subequations}\label{eq:opf}
\begin{align}
\min_{\substack{\boldsymbol{\widecheck{v}},\boldsymbol{\widecheck{\theta}}, \\ \boldsymbol{\widecheck{p}}_{\mathrm{g}}, \boldsymbol{\widecheck{q}}_{\mathrm{g}}, \\ \boldsymbol{p}_{\mathrm{dg},j}, \\ \boldsymbol{q}_{\mathrm{dg},j}}} \quad & \sum_{i = 1}^{n_{\mathrm{g}}}C_{i}(\boldsymbol{{\widecheck{p}}}_{\mathrm{g}}^{(i)}) + \sum_{j = 1}^{n_{\mathrm{ds}}}\sum_{k = 1}^{n_{\mathrm{dg},j}}C_{jk}(\boldsymbol{p}_{\mathrm{dg},j}^{(k)}) \label{eq:opf0}\\
\text{s.t.} \quad 
& {G}_{\mathrm{P}}(\boldsymbol{\widecheck{v}},  \boldsymbol{\widecheck{\theta}}; \widecheck{Y}) + \widecheck{p}_{\mathrm{d}} - \boldsymbol{\widecheck{p}}_{\mathrm{g}} = 0, \label{eq:opf1} \\
& {G}_{\mathrm{Q}}(\boldsymbol{\widecheck{v}},  \boldsymbol{\widecheck{\theta}}; \widecheck{Y}) + \widecheck{q}_{\mathrm{d}} - \boldsymbol{\widecheck{q}}_{\mathrm{g}} = 0, \label{eq:opf2} \\
& G_{\mathrm{line}}(\boldsymbol{\widecheck{v}},  \boldsymbol{\widecheck{\theta}}; \widecheck{Y}) \preceq \widecheck{l}_{\mathrm{line, max}}, \label{eq:opf3} \\
& \widecheck{v}_{\mathrm{min}} \preceq \boldsymbol{\widecheck{v}} \preceq \widecheck{v}_{\mathrm{max}}, \ \widecheck{\theta}_{\mathrm{min}} \preceq \boldsymbol{\widecheck{\theta}} \preceq \widecheck{\theta}_{\mathrm{max}}, \label{eq:opf4} \\
& \widecheck{p}_{\mathrm{g, min}} \preceq \boldsymbol{\widecheck{p}}_{\mathrm{g}} \preceq \widecheck{p}_{\mathrm{g, max}}, \ \widecheck{q}_{\mathrm{g, min}} \preceq \boldsymbol{\widecheck{q}}_{\mathrm{g}} \preceq \widecheck{q}_{\mathrm{g, max}}, \label{eq:opf5}\\
& P_{j, u}(\boldsymbol{x}_{j}) + \boldsymbol{\widecheck{p}}_{\mathrm{g}}^{(s_{j, u})} = 0, \ \forall j \in \{1,.., n_{\mathrm{ds}}\}, \notag \\
& \hspace{3.55cm} \forall u \in \{1,.., r_{j} \}, \label{eq:opf6} \\
& Q_{j, u}(\boldsymbol{x}_{j}) + \boldsymbol{\widecheck{q}}_{\mathrm{g}}^{(s_{j, u})} = 0, \ \forall j \in \{1,.., n_{\mathrm{ds}}\}, \notag \\
& \hspace{3.6cm} \forall u \in \{1,.., r_{j}\}, \label{eq:opf7} \\
& FR_{j}(\boldsymbol{x}_{j}) \preceq 0, \ \forall j \in \{1,.., n_{\mathrm{ds}}\}, \label{eq:opf8}\\
& x_{j, \mathrm{min}} \preceq \boldsymbol{x}_{j} \preceq x_{j, \mathrm{max}}, \ \forall j \in \{1,.., n_{\mathrm{ds}}\}, \label{eq:opf9}\\
& \boldsymbol{x}_{j} = {\begin{bmatrix}
 \boldsymbol{v}_{j}^{\top} \ {\boldsymbol{p}_{\mathrm{dg},j}}^{\top} \  {\boldsymbol{q}_{\mathrm{dg},j}}^{\top}\end{bmatrix}}^{\top}, \forall j \in \{1,.., n_{\mathrm{ds}}\}. \label{eq:opf10}
\end{align}
\end{subequations}

Examining Equations \eqref{eq:opf1} - \eqref{eq:opf5}, it can be seen that these equations contain only TS-related variables. The DS-related variables are expressed through the functions defined in Equations \eqref{eq:opf6} - \eqref{eq:opf8}. The functions $P_{j, u}(\boldsymbol{x}_{j})$ and $Q_{j, u}(\boldsymbol{x}_{j})$ are designed to represent DS-related variables to the active and reactive power flow at the PCCs. Thanks to these functions, DSs are modeled as active and reactive power sources at the PCCs from the perspective of the TS. Note that the variables $\boldsymbol{\widecheck{p}}_{\mathrm{g}}^{(s_{j, u})}$ and $\boldsymbol{\widecheck{q}}_{\mathrm{g}}^{(s_{j, u})}$ represent the active and reactive power flow at the PCC, respectively, directed from DS towards TS.

The functions $FR_{j}(\boldsymbol{x}_{j})$ are designed to represent the feasible region of the DSs. These functions ensure that technical constraints, such as line flow and voltage magnitude limits within the DS, are satisfied. Specifically, if $\boldsymbol{x}_{j}$ represents a feasible operating point that complies with all DS constraints, the condition $FR_{j}(\boldsymbol{x}_{j}) \preceq 0$ is satisfied. If this condition is not met, it indicates that $\boldsymbol{x}_{j}$ lies outside the feasible region. Moreover, \eqref{eq:opf9} defines the bounds for the DS-related variables. It should be noted that for each DS, only a single $FR_{j}(\boldsymbol{x}_{j})$ function is created, regardless of the number of PCCs. However, for each DS, separate $P_{j, u}(\boldsymbol{x}_{j})$ and $Q_{j, u}(\boldsymbol{x}_{j})$ functions must be defined for each PCC.

In summary, we encapsulate non-sensitive DS-related variables within a specific set of functions to represent the technical constraints of the DS while preserving data privacy. These functions are constructed using ML models trained exclusively on non-sensitive DS-related variables, ensuring that sensitive data remains protected throughout the process.

\begin{figure*}
\centering
\includegraphics[width=0.7\textwidth]{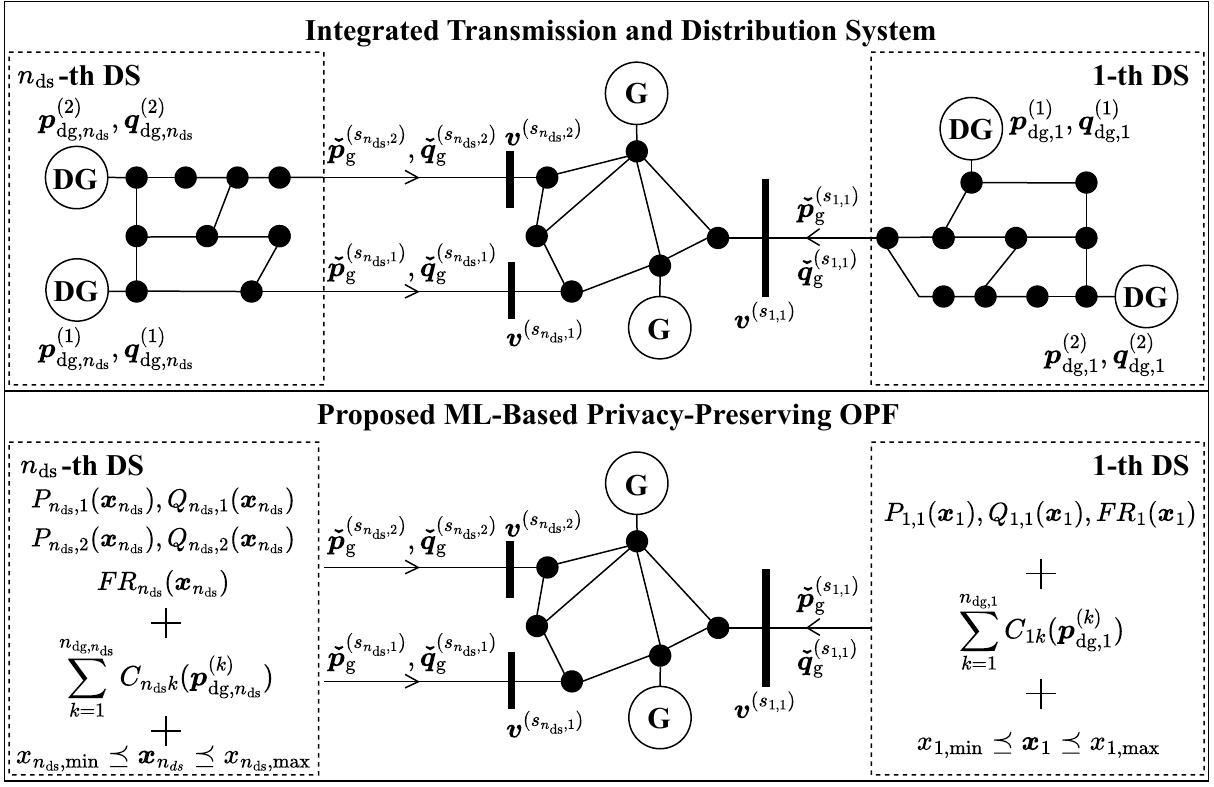}
\caption{Schematic representation of the proposed method.}
\label{fig:method_multi_pcc}
\end{figure*}

Fig. \ref{fig:method_multi_pcc} illustrates the schematic representation of the proposed method. As outlined in previous sections, the OPF should be solved by TSOs. To facilitate this, the ML models and the cost functions of the DGs are shared with the TSO. By employing these ML models, the TSO can effectively solve the proposed ML-based privacy-preserving OPF \eqref{eq:opf}. This approach allows the OPF to be solved and the dispatch decisions for the DGs to be determined in a single round of communication, without requiring any additional disaggregation processes. Consequently, the flexibility obtained from DSs can be utilized for various network management purposes in a cost-effective manner, while ensuring the protection of sensitive data and adhering to the technical constraints of both TSOs and DSOs.

\section{Dataset Creation}\label{sec:dataset_creation}

In the proposed method, we represent the technical constraints of the DSs using a set of functions developed through ML models. The effective training of these models necessitates a comprehensive dataset. To generate this dataset, we employ the Latin Hypercube Sampling (LHS) method \cite{huntington1998improvements}. LHS allows for sampling from a multidimensional distribution while maintaining the marginal probability distributions for each variable. This technique ensures efficient exploration of the entire range of each variable, even when the number of samples is relatively small.

To create the dataset, the DSO generates various operating points, represented by different values of $\boldsymbol{x}_{j}$, within the specified limits of these variables, as outlined in \eqref{eq:opf9}. Particular attention must be given to the variables $\boldsymbol{p}_{\mathrm{dg},j}$ and $\boldsymbol{q}_{\mathrm{dg},j}$, as they define the PQ chart of the flexibility-providing units (FPUs). It is important to note that the DGs used in present study can be also considered as FPUs.

\begin{figure}
\centering
\includegraphics[width=\columnwidth]{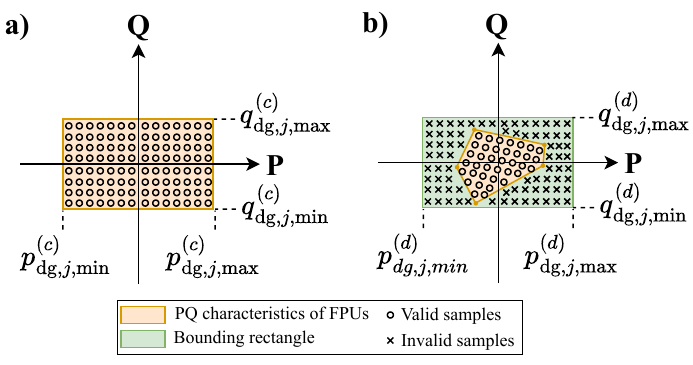}
\caption{Data sampling approach using LHS.}
\label{fig:pq_chart}
\end{figure}

In most studies, the PQ characteristics of FPUs are typically considered as rectangular (ideal or generic) \cite{schneider2021estimation} (see Fig. \ref{fig:pq_chart}a). However, FPUs exhibit varying PQ characteristics, which are often modeled as convex polygons \cite{riaz2019feasibility}. In \cite{contreras2021computing}, rather than focusing on specific FPU shapes, such as triangular or square configurations, the methodology is demonstrated using arbitrary convex polygons. This approach illustrated the applicability of the method across diverse characteristics. Following this direction, we also represent PQ characteristics using randomly generated arbitrary convex polygons, thereby demonstrating the effectiveness of the proposed method for different FPU characteristics (see Fig. \ref{fig:pq_chart}b).

In generating the dataset, we introduce a novel approach for sampling with LHS in scenarios involving randomly generated arbitrary convex polygons. In this approach, $p_{\mathrm{dg},j, \mathrm{min}}$, $p_{\mathrm{dg},j, \mathrm{max}}$, $q_{\mathrm{dg},j, \mathrm{min}}$ and $q_{\mathrm{dg},j, \mathrm{max}}$ are determined such that the rectangles formed by these values fully encapsulate the arbitrary convex polygons. Subsequently, LHS is applied within these bounding rectangles, enabling sampling from the entire arbitrary convex polygon that lies within the bounding rectangle. 

It is important to note that, as a natural consequence of this approach, some samples are taken from the area between the arbitrary polygon and the bounding rectangle. However, these samples do not represent feasible operating points. Fig. \ref{fig:pq_chart} illustrates the data sampling approach using LHS. Specifically, Fig. \ref{fig:pq_chart}a illustrates a rectangular PQ characteristic, while Fig. \ref{fig:pq_chart}b shows a PQ characteristic of a convex polygon along with its bounding rectangle. It also distinguishes between valid samples that fall within the convex polygon and invalid samples that are located in the region between the polygon and the bounding rectangle.

To accurately use only valid samples within the polygon, another approach is required. As is well known, convex polygons can be characterized by a set of linear inequalities. Accordingly, we define the linear inequalities as follows:

\begin{equation}\label{eq:polygon}
   \resizebox{\columnwidth}{!}{$
   A_{\mathrm{PQ},jk} \ {\begin{bmatrix} 
 \boldsymbol{p}_{\mathrm{dg},j}^{(k)}  &\boldsymbol{q}_{\mathrm{dg},j}^{(k)}\end{bmatrix}}^{\top} \preceq b_{\mathrm{PQ},jk} \forall j \in \{1,.., n_{\mathrm{ds}}\} \text{ and } \forall k \in \{1,.., n_{\mathrm{dg},j}\},
   $}
\end{equation}

\noindent where $A_{\mathrm{PQ},jk} \in \mathbb{R}^{{n}_{\mathrm{v},jk} \times 2}$ represents the matrix of coefficients, while $b_{\mathrm{PQ},jk} \in \mathbb{R}^{{n}_{\mathrm{v},jk}}$ is the vector of constants. Also, ${n}_{\mathrm{v},jk}$ indicates the number of vertices that define the convex polygon for a given DG. Note that, for DGs with rectangular characteristics, the vertices are determined by the $p_{\mathrm{dg},j, \mathrm{min}}$, $p_{\mathrm{dg},j, \mathrm{max}}$, $q_{\mathrm{dg},j, \mathrm{min}}$ and $q_{\mathrm{dg},j, \mathrm{max}}$ values. After defining these linear inequalities, they can be incorporated into the OPF problem defined in \eqref{eq:opf} to ensure that invalid samples, which lie between the convex polygon and the bounding rectangle, are identified as infeasible. To achieve this, we can extend the OPF problem as follows:

\begin{equation}\label{eq:opf_polygon}
\begin{aligned}
\min \quad & \eqref{eq:opf0} \\
\text{s.t.} \quad & \eqref{eq:opf1} - \eqref{eq:opf10}, \\
& \eqref{eq:polygon}.
\end{aligned}
\end{equation}

This ensures that invalid samples are appropriately classified as infeasible, allowing only valid samples to be evaluated. This approach facilitates dataset generation using standard LHS without the need for additional sampling techniques. Consequently, the proposed method can effectively handle FPUs with arbitrary convex polygon characteristics beyond rectangular ones. Furthermore, since the FPUs characteristics are represented by linear inequalities, they can be integrated into the OPF  problem in a computationally efficient manner.

Overall, each operating point $\boldsymbol{x}_{j}$ generated by LHS is assessed based on the security limits of the DSs using power flow analysis. Based on this evaluation, the operating points are classified as either feasible or infeasible. Subsequently, datasets are compiled consisting of feasible instances \(\mathcal{F}\) and infeasible instances \(\mathcal{I}\).

\section{Representation Technical Constraints of the Distribution Systems with Machine Learning Models}\label{sec:ml_models}

After creating the dataset, the ML models are trained on this data to generate the previously defined functions. Specifically, quadratic regression models are employed to construct $P_{j, u}(\boldsymbol{x}_{j})$ and $Q_{j, u}(\boldsymbol{x}_{j})$ functions. In addition to these, the $FR_{j}(\boldsymbol{x}_{j})$ functions, which are designed to represent the feasible region of the DSs, are implemented as classification models. To accomplish this, we introduce a novel, tailored NN model.

\subsection{NN-Guided Polytope Representation of Feasible Region}\label{sec:classification}

In this section, we model the functions $FR_{j}(\boldsymbol{x}_{j})$ in the form of a convex polytope, ensuring that the condition $FR_{j}(\boldsymbol{x}_{j}) \preceq 0$ is satisfied. To construct this model, we utilize a previously generated dataset that consists of a finite number of feasible and infeasible instances i.e., \(\mathcal{F}\) and \(\mathcal{I}\). Following this, we can describe the function as follows: 

\begin{equation}\label{eq:fr}
   FR_{j}(\boldsymbol{x}_{j}) = A_{\mathrm{FR},j} \boldsymbol{x}_{j} - b_{\mathrm{FR},j}. 
\end{equation}

To construct a convex polytope that encompasses all feasible instances, we need to determine a matrix \(A_{\mathrm{FR},j} \in \mathbb{R}^{n_{\mathrm{f},j} \times n_{j}}\) and a vector \(b_{\mathrm{FR},j} \in \mathbb{R}^{n_{\mathrm{f},j}}\). This formulation ensures that all feasible instances \(\boldsymbol{x}_{j} \in \mathcal{F}\) satisfy the inequality \(A_{\mathrm{FR}, j}\boldsymbol{x}_{j} \preceq b_{\mathrm{FR},j}\), while all infeasible instances \(\boldsymbol{x}_{j} \in \mathcal{I}\) do not satisfy this inequality, i.e., \(A_{\mathrm{FR},j}\boldsymbol{x}_{j} \not\preceq b_{\mathrm{FR},j}\). Note that, \(n_{\mathrm{f},j}\) represents the number of facets of the polytope, assuming that there is no redundancy in \(A_{\mathrm{FR}, j}\boldsymbol{x}_{j} \preceq b_{\mathrm{FR},j}\).

\begin{figure}
\centering
\includegraphics[width=\columnwidth]{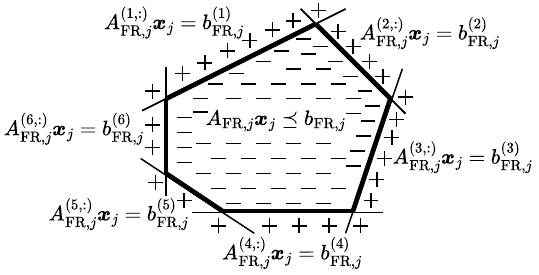}
\caption{Schematic representation of the polytope.}
\label{fig:polytope_multi}
\end{figure}

The next step in constructing the polytope involves determining under what circumstances an operating point \(x_{j}\) satisfies the defined inequality. We consider \(x_{j}\) to satisfy the inequality if and only if every element of $z$ is less than or equal to zero, where $z = A_{\mathrm{FR},j}x_{j} - b_{\mathrm{FR},j}$. As a result, \(\max(z) \leq 0\) indicates that \(x_{j}\) lies inside the polytope, making it a feasible point. On the contrary, if at least one element of \(x_{j}\) is strictly greater than zero, this implies $\max{(z)} > 0$, meaning that \(x_{j}\) is outside the polytope and therefore an infeasible instance. To better understand this polytope, Fig. \ref{fig:polytope_multi} provides a schematic representation with \(n_{\mathrm{f},j} = 6\), where feasible instances are depicted by $-$ and and infeasible instances by $+$.

After determining the approach to assess whether a given operating point \(x_{j}\) lies inside or outside the polytope, the next crucial step is to define the appropriate parameters \(A_{\mathrm{FR},j}\) and \(b_{\mathrm{FR},j}\). To achieve this, we leverage a novel tailored NN architecture specifically designed for this purpose. Upon training, the weights and biases of this NN model are directly mapped to the parameters \(A_{\mathrm{FR},j}\) and \(b_{\mathrm{FR},j}\). In this framework, feasible instances are labeled as Class 0, and infeasible instances as Class 1. The proposed NN architecture can be mathematically represented as follows:

\begin{subequations}\label{eq:nnet}
\begin{align}
& \boldsymbol{o}_{j} = W_{j}\boldsymbol{x}_{j} + b_{j}, \label{eq:nnet1} \\
& \boldsymbol{f}_{j} = \max(\boldsymbol{o}_{j}),  \label{eq:nnet2} \\
& \boldsymbol{y}_{j} = \mathrm{sigmoid}(\boldsymbol{f}_{j}). \label{eq:nnet3}
\end{align}
\end{subequations}

Equation \eqref{eq:nnet} describes a feed-forward architecture. Firstly, \eqref{eq:nnet1} represents a hidden layer with \(n_{h,j}\) nodes, where \(W_{j} \in \mathbb{R}^{n_{\mathrm{h},j} \times n_{j}}\) denotes the weight matrix and \(b_{j} \in \mathbb{R}^{n_{\mathrm{h},j}}\) denotes the bias vector. In \eqref{eq:nnet2}, instead of using a standard activation function, the output of the hidden layer \(\boldsymbol{o}_{j} \in \mathbb{R}^{n_{\mathrm{h},j}}\) is processed by a max aggregator function, resulting in \(\boldsymbol{f}_{j} \in \mathbb{R}\). Then, \(\boldsymbol{f}_{j}\) is passed through the sigmoid activation in \eqref{eq:nnet3}, producing the final output \(\boldsymbol{y}_{j} \in [0,1]\), which can be interpreted as the probability of infeasibility of a given input \(\boldsymbol{x}_{j} \). Fig. \ref{fig:nn_architecture_multi} shows the architecture of the proposed NN. Finally, to train the NN, we employ the standard binary cross-entropy loss function. Additionally, we introduce weights (penalties) within the loss function $L_{j}(\cdot)$ as follows:

\begin{figure}
\centering
\includegraphics[width=\columnwidth]{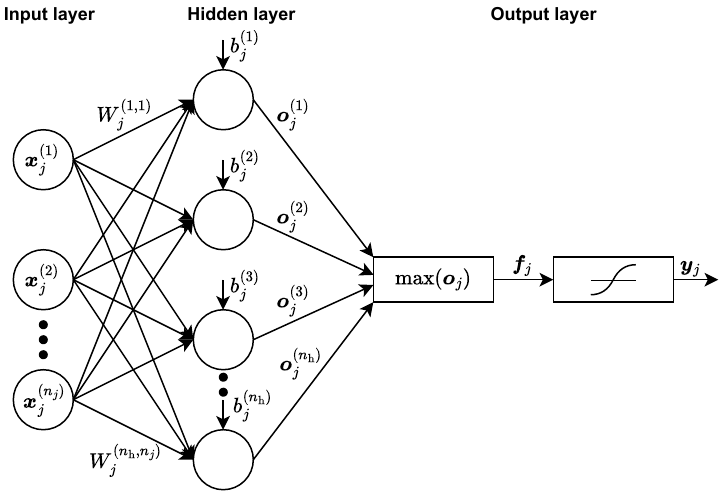}
\caption{The architecture of the novel tailored NN.}
\label{fig:nn_architecture_multi}
\end{figure}

\begin{align}\label{eq:nnet-class}
L_{j} = -\sum_{i}w_{j,10} \,  y_{j,i} \,  \log(\widetilde{y}_{j,i}) + w_{j,01} \,  (1 - y_{j,i}) \,  \log(1 - \widetilde{y}_{j,i}), 
\end{align}

\noindent where $y_{j,i} \in \{0,1\}$ represents the true output for the instance $i$, while $\widetilde{y}_{j,i} \in [0,1]$ is the predicted output for the same instance. The weight $w_{j,10} > 0$ is assigned to penalize the incorrect classification of an infeasible instance as feasible, while $w_{j,01} > 0$ represents vice versa. The use of different weights is crucial because incorrect classifications have different consequences. Incorrectly classifying a feasible point as infeasible may only lead to economic losses, while misclassifying an infeasible point as feasible can result in significant issues within the power system. To address this, we assign higher weights to the misclassification of infeasible points as feasible (i.e., $w_{10}$). This approach helps prevent the NN from classifying an infeasible point as feasible. Furthermore, by adopting this strategy, we can represent the non-convex feasible area as a conservative convex polytope. While this approximation may introduce a slight increase in the total cost, it ensures a more reliable representation of the feasible region. 

As the final step, the relationship between the weights \(W_{j}\) and biases \(b_{j}\) of the NN and matrix \(A_{\mathrm{FR},j}\) and vector \(b_{\mathrm{FR},j} \) needs to be established. According to \eqref{eq:nnet}, a sample \(\boldsymbol{x}_{j}\) is classified as feasible if \(\max(W_{j}\boldsymbol{x}_{j}+b_{j}) \leq 0\) and as infeasible if \(\max(W_{j}\boldsymbol{x}_{j}+b_{j}) > 0\). This implies that the decision region for feasible samples is defined by $W_{j}\boldsymbol{x}_{j} \preceq -b_{j}$. Thus, the desired polytope can be defined by setting \(A_{\mathrm{FR},j} = W_{j}\) and \(b_{\mathrm{FR},j} = -b_{j}\). Note that, the number of hidden nodes \(n_{\mathrm{h},j}\) provides an upper bound on the number of facets of the polytope (though this is only an upper bound, as some rows of \(W_{j}\boldsymbol{x}_{j} \preceq -b_{j}\) may be redundant). Consequently, by incorporating the constraint $A_{\mathrm{FR},j}\boldsymbol{x}_{j} \preceq b_{\mathrm{FR},j}$ into the OPF problem as specified \eqref{eq:opf8}, the feasible region of the DS can be effectively approximated. Utilizing such a polytope allows the OPF to be implemented in a computationally efficient and privacy-preserving manner.

\subsection{Quadratic Regression-Based Power Flow Approximator}\label{regression}

After defining the feasible region of the DSs, we focus on defining the functions $P_{j, u}(\boldsymbol{x}_{j})$ and $Q_{j, u}(\boldsymbol{x}_{j})$. These functions are designed to map the DS-related variables $\boldsymbol{x}_{j}$ to the active and reactive power flows at the PCCs, respectively. Considering the inherent quadratic relationship between power injections and system losses \cite{liu2018estimating}, we select a quadratic regression model to define these mappings. Accordingly, these functions can be described as follows:

\begin{subequations}\label{eq:pq}
\begin{align}
& P_{j,u}(\boldsymbol{x}_{j}) = \boldsymbol{x}_{j}^{\top} A_{\mathrm{P},j,u} \boldsymbol{x}_{j} + b_{\mathrm{P},j,u}^{\top}\boldsymbol{x}_{j} + c_{\mathrm{P},j,u}, \\
& Q_{j,u}(\boldsymbol{x}_{j}) = \boldsymbol{x}_{j}^{\top} A_{\mathrm{Q},j,u} \boldsymbol{x}_{j} + b_{\mathrm{Q},j,u}^{\top}\boldsymbol{x}_{j} + c_{\mathrm{Q},j,u},
\end{align}
\end{subequations}

\noindent where $A_{\mathrm{P},j,u}, A_{\mathrm{Q},j,u}  \in \mathbb{R}^{n_{\mathrm{j}} \times n_{\mathrm{j}}}$, and $b_{\mathrm{P},j,u}, b_{\mathrm{Q},j,u} \in \mathbb{R}^{n_{\mathrm{j}}}$, and $c_{\mathrm{P},j,u}, c_{\mathrm{Q},j,u} \in \mathbb{R}$ represent the model parameters, which are determined through standard ML training procedures.

To train the ML models, we use $\boldsymbol{x}_{j}$ as input, active and reactive power at the PCCs, i.e., $\boldsymbol{\widecheck{p}}_{\mathrm{g}}^{(s_{j, u})}$ and $\boldsymbol{\widecheck{q}}_{\mathrm{g}}^{(s_{j, u})}$ as output of the models. The values of $\boldsymbol{\widecheck{p}}_{\mathrm{g}}^{(s_{j, u})}$ and $\boldsymbol{\widecheck{q}}_{\mathrm{g}}^{(s_{j, u})}$ are derived from power flow calculations. It is important to note that only feasible instances \(\mathcal{F}\) are utilized during the training process, ensuring that the model learns the mapping from operating points that are feasible. Upon completion of the training process, these functions accurately represent the relationship between DG-related variables and the corresponding power flows at the PCCs.

\section{Case Studies and Discussion}\label{sec:case}

In the present paper, the performance of the proposed method is evaluated by comparing it with the traditional AC-OPF, which does not consider data privacy. The evaluation process involves several steps: first, a suitable power system is established, followed by the generation of a comprehensive dataset. ML models are then trained using this dataset, and their approximation accuracy is assessed. Finally, the effectiveness of the proposed method are examined through extensive case studies.

The primary simulation environment for this study is \textsc{MATLAB}, where we utilize \textsc{MATPOWER} \cite{MATPOWER} with \textsc{KNITRO} solver \cite{byrd2006k} for performing AC-OPF calculations. ML models are developed and trained using \textsc{TensorFlow}/\textsc{Keras} \cite{tensorflow2015-whitepaper, chollet2015keras}. The case studies are conducted on a PC equipped with an Intel Core i7-10700K CPU @ 3.80 GHz and 32 GB RAM.

\subsection{Power System Creation}

\begin{figure}
\centering
\includegraphics[width=\columnwidth]{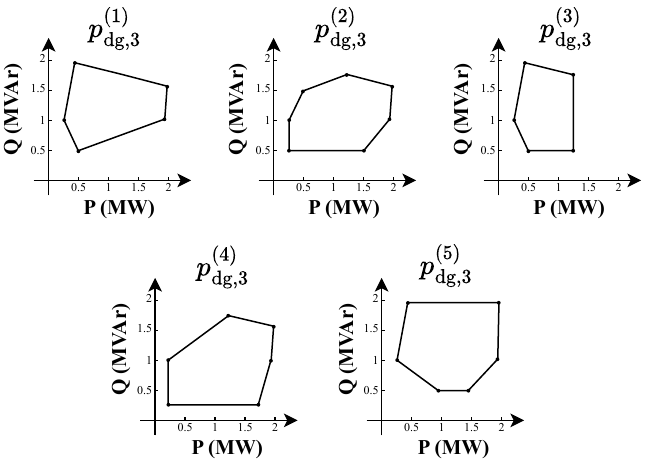}
\caption{Convex polygon PQ characteristics for DGs in the third DS.}
\label{fig:pq_chart_case_study}
\end{figure}

To create an appropriate integrated transmission and distribution system, we employ IEEE 30-bus test system as TS, and three IEEE 33-bus test systems as the DSs. Fig. \ref{fig:pq_chart_case_study} shows the integrated power system configuration. It is important to note that a relatively small TS is deliberately chosen to enable a more rigorous comparison of the proposed method under challenging conditions. Otherwise, in an integrated power system with a large TS, the impact of the DSs would be minimal, making it difficult to accurately assess the effectiveness of the proposed method. 

For all DSs, the normally open lines are closed, converting the systems into meshed grids to assess the effectiveness of the proposed method in meshed grids. The first DS is connected to the 11th bus of the TS (i.e., $s_{1,1}=11$) through a single PCC, and includes only one DG. This setup is designed such that the first DS is represented in a three-dimensional space (i.e., $\boldsymbol{x}_{1} \in \mathbb{R}^{3}$), enabling the proposed method to be visualized within three-dimension.

\begin{figure}
\centering
\includegraphics[width=\columnwidth]{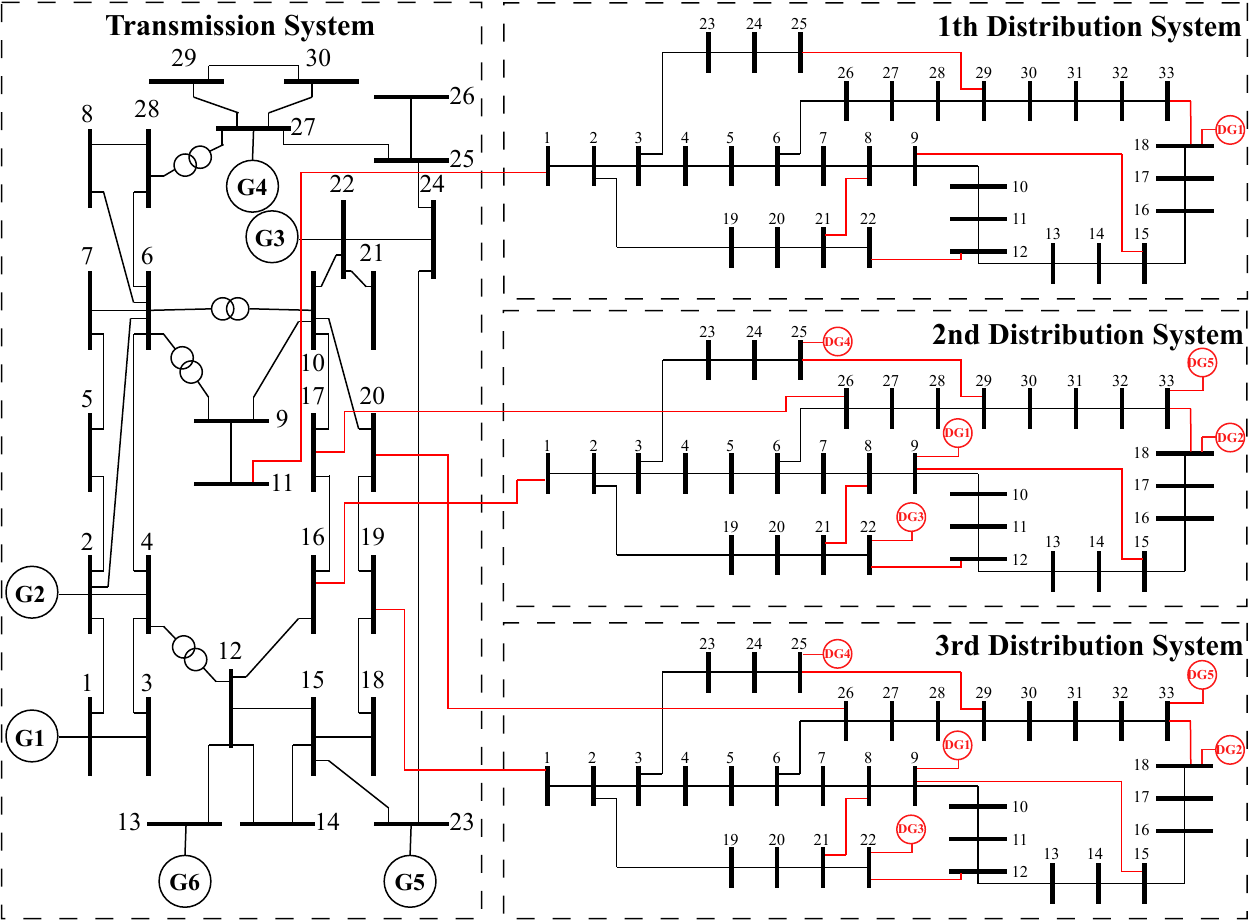}
\caption{Single line diagram of the integrated power system.}
\label{fig:tso_dso}
\end{figure}

To demonstrate the effectiveness of the proposed method in a more complex system, five DGs are integrated into the second DS. Additionally, as is commonly observed in real-world scenarios, this DS is connected to the TS through two PCCs located at the 16th and 17th buses of the TS (i.e., $s_{2,1}=16$ and $s_{2,2}=17$), effectively showcasing the scenario involving multiple PCCs. The third DS demonstrates the effectiveness of the proposed method on FPUs with varying PQ characteristics. To achieve this, the same topology as the second DS is employed, but instead of DGs with only rectangular PQ charts as in the first two DSs, this system incorporates DGs with varying convex polygon PQ characteristics, as illustrated in Fig. \ref{fig:pq_chart_case_study}. This setup demonstrates that the proposed method maintains high performance regardless of the specific PQ characteristics. Additionally, the third DS is connected to the TS through two PCCs  at the 19th and 20th buses of the TS (i.e., $s_{3,1}=19$ and $s_{3,2}=20$). Furthermore, all DGs are designed to have active power outputs ranging between 0 and 2 MW, and reactive power outputs between 0 and 2 MVAr, respectively. Note that these values also define the bounding rectangle for the DGs within the third DS.

\subsection{Dataset Generation, Training, and Approximation Quality in ML Models}

To develop the ML models, the first step involves generating the dataset, for which we employ LHS, as detailed earlier. For the first DS, a total of 20,000 data points are generated, while 500,000 data points are generated for both the second and third DSs. It is important to note that, as previously discussed, the same dataset and ML models are used for both the second and third DSs to demonstrate that the proposed method maintains high performance, irrespective of the specific PQ characteristics. After generating the datasets, we train our ML models by following standard procedures for both classification and regression tasks. This includes separating the dataset into a training set (80\%) and a test set (20\%) to validate the model's performance.

We employ NN classification models to distinguish between feasible and infeasible operating points based on the generated $\boldsymbol{x}_{j}$ values. For constructing these NN models, we employ random search hyperparameter tuning, selecting $n_{\mathrm{h},1} = 20$ hidden nodes for the first DS, and $n_{\mathrm{h},2} = 1,000$ and $n_{\mathrm{h},3} = 1,000$ hidden nodes for the second and third DSs, respectively. As the complexity of the power system increases, the number of facets required to accurately represent the feasible space of the DSs also rises, hence the increased number of hidden nodes. It is important to recall that the number of hidden nodes, $n_{\mathrm{h},j}$,  provides an upper bound on the number of facets that define the polytope. Consequently, some rows of the matrix $W_{j}$ and vector $b_{j}$ may be redundant. Therefore, even when a large number of hidden nodes are defined, the NN only generates as many facets as necessary to describe the feasible region effectively. 

Furthermore, for the second and third DSs, we select $w_{2,10}$ and $w_{3,10}$ as 2, and $w_{2,01}$ and $w_{2,01}$ as 1. For the first DS, both $w_{1,10}$ and $w_{1,01}$ are set to 1. These weights are chosen to account for the increased complexity of the feasible space in the second and third DSs. In more complex models, the NN needs to be more conservative in defining the feasible space, which helps to avoid the misclassification of infeasible points as feasible. Such misclassification could lead to significant operational issues in the power system, thus justifying the more cautious approach in these cases.

\begin{figure}
\centering
\includegraphics[width=\columnwidth]{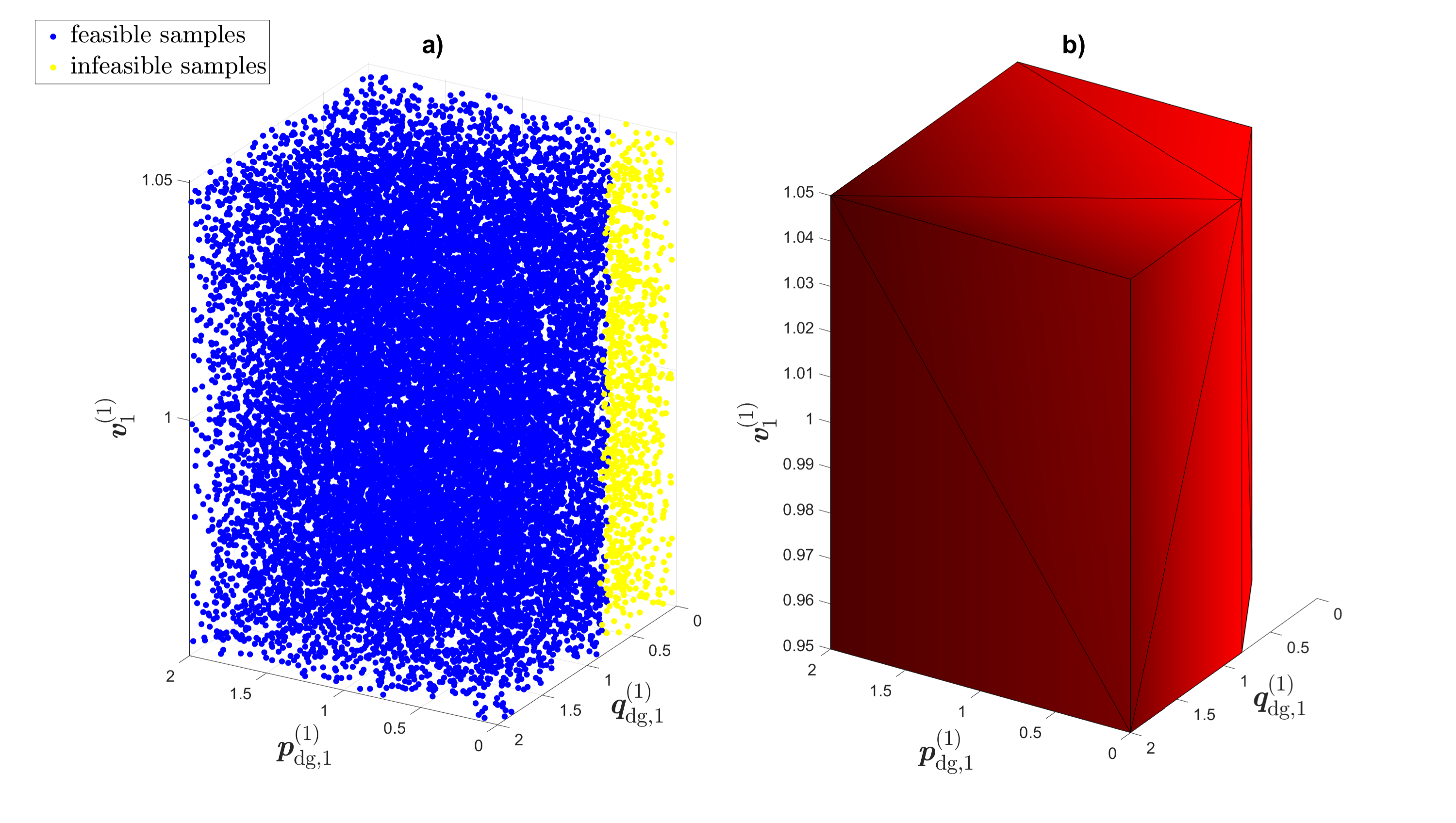}
\caption{a) Dataset indicating feasible and infeasible samples. b) Feasible region approximation of the NN indicating the facets.}
\label{fig:dataset_nn}
\end{figure}

After training the models, we assess their approximation quality by evaluating them on the test sets using accuracy, recall, and specificity metrics \cite{hossin2015review}. The results are summarized in Table \ref{tab:classification_metric}. For the NN model of the first DS, i.e., $FR_{1}(\boldsymbol{x}_{1})$, which is a relatively less complex system, the all metrics are observed to be nearly 100\%. Additionally, Fig. \ref{fig:dataset_nn} provides a visual representation of the generated dataset and the NN's approximation of the feasible region for the first DS. Notably, if the voltage were assumed to be constant, the feasible region would be represented as a two-dimensional area. However, the figure illustrates a larger three-dimensional region, demonstrating that incorporating voltage variations allows for better utilization of DS flexibility potential. As depicted in Fig. \ref{fig:dataset_nn}, the NN model accurately approximates the feasible region and successfully establishes a well-defined decision boundary.

\begin{table}
\centering
\caption{Accuracy, Recall and Specificity Metrics of the NN Models}
\label{tab:classification_metric}
\resizebox{\columnwidth}{!}{%
\begin{tabular}{lccc}
\hline
\multicolumn{1}{l}{\multirow{2}{*}{Model}} &
  \multicolumn{1}{c}{\multirow{2}{*}{Accuracy}} &
  \multicolumn{1}{c}{\multirow{2}{*}{Recall}} &
  \multicolumn{1}{c}{\multirow{2}{*}{Specificity}} \\
\multicolumn{1}{l}{}    & \multicolumn{1}{c}{}       & \multicolumn{1}{c}{}       & \multicolumn{1}{c}{}       \\ \hline
$FR_{1}(\boldsymbol{x}_{1})$                       & 99.90\%                      & 99.89\%                      & 100.00\%                      \\
\multicolumn{1}{l}{$FR_{2}(\boldsymbol{x}_{2})$ - $FR_{3}(\boldsymbol{x}_{3})$} & \multicolumn{1}{c}{94.79\%} & \multicolumn{1}{c}{93.03\%} & \multicolumn{1}{c}{97.01\%} \\ \hline
\end{tabular}%
}
\end{table}

When examining the models for the second and third DS, i.e., $FR_{2}(\boldsymbol{x}_{2})$ and $FR_{3}(\boldsymbol{x}_{3})$, it is observed that the accuracy and recall metrics are 94.79\% and 93.03\%, respectively. The slightly lower values are attributed to the weights (penalties) applied during the training process. Specifically, the model is penalized more heavily for predicting an infeasible point as feasible, which results in slightly lower accuracy and recall, as the model becomes biased towards predicting data points as infeasible. Correspondingly, the specificity metric is 97.01\%, indicating that the model is highly effective at avoiding misclassification of infeasible points as feasible, as desired. This approach, which prioritizes preventing of predicting infeasible operating points as feasible, naturally leads to a slight trade-off in accuracy. However, this ensures that the NN models effectively prevent results at infeasible operating points, accepting minor economic losses in favor of operational security.

Following the NN classification models, we develop quadratic regression models and evaluate their performance using numerical metrics such as root mean square error (RMSE) and mean absolute error (MAE).  The results, displayed in Table \ref{tab:regression_metric}, indicate that all regression models achieved performances close to 100\%. These numerical results clearly demonstrate that the ML models are highly effective in capturing and mapping the characteristics of the DSs.

\begin{table}
\centering
\caption{RMSE and MAE Metrics of the Quadratic Regression Models}
\label{tab:regression_metric}
\resizebox{0.8\columnwidth}{!}{%
\begin{tabular}{lcc}
\hline
\multirow{2}{*}{Model} & \multirow{2}{*}{RMSE} & \multirow{2}{*}{MAE} \\
                       &                       &                      \\ \hline
$P_{1,1}(\boldsymbol{x}_{1})$                   & 5.0$\times$10\textsuperscript{-4}                    & 3.6$\times$10\textsuperscript{-4}                   \\ 
$P_{2,1}(\boldsymbol{x}_{2})$ - $P_{3,1}(\boldsymbol{x}_{3})$            & 2.9$\times$10\textsuperscript{-4}                    & 2.0$\times$10\textsuperscript{-4}                   \\ 
$P_{2,2}(\boldsymbol{x}_{2})$ - $P_{3,2}(\boldsymbol{x}_{3})$            & 4.9$\times$10\textsuperscript{-4}                    & 3.4$\times$10\textsuperscript{-4}                   \\ 
$Q_{1,1}(\boldsymbol{x}_{1})$                  & 4.3$\times$10\textsuperscript{-4}                    & 3.0$\times$10\textsuperscript{-4}                 \\ 
$Q_{2,1}(\boldsymbol{x}_{2})$ - $Q_{3,1}(\boldsymbol{x}_{3})$            & 2.7$\times$10\textsuperscript{-4}                    & 1.8$\times$10\textsuperscript{-4}                   \\ 
$Q_{2,2}(\boldsymbol{x}_{2})$ - $Q_{3,2}(\boldsymbol{x}_{3})$            & 4.5$\times$10\textsuperscript{-4}                    & 3.1$\times$10\textsuperscript{-4}                   \\ \hline
\end{tabular}%
}
\end{table}

\subsection{Benchmark Results}

In the present section, we perform a comprehensive analysis by comparing the proposed method with standard AC-OPF across 1,000 randomly generated sets of cost coefficients, focusing on total cost and computational time. This approach enables a detailed evaluation of the proposed method's effectiveness. Notably, to assess the method's performance on FPUs with diverse PQ characteristics, we use the convex polygon characteristics illustrated in Fig. \ref{fig:pq_chart_case_study}. These characteristics are defined as sets of linear inequalities, as given in \eqref{eq:polygon}, and then the proposed method is applied via \eqref{eq:opf_polygon}. The results are subsequently compared with those of the standard AC-OPF approach, formulated in \eqref{eq:opf_dsotso}.

\begin{figure}
\centering
\includegraphics[width=\columnwidth]{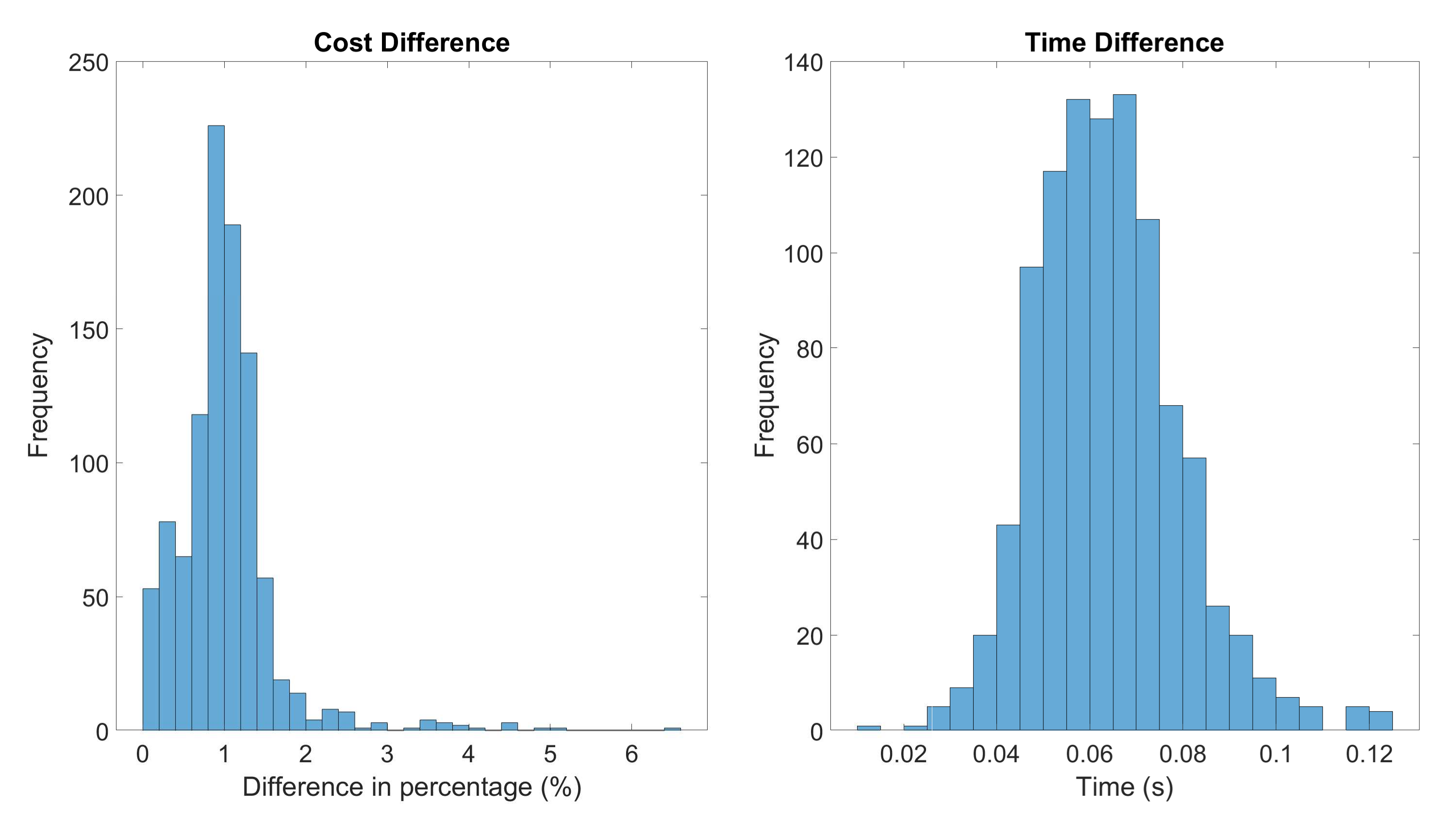}
\caption{The histogram of the total cost and computational time differences taking
AC-OPF as reference.}
\label{fig:histogram}
\end{figure}

The histogram in Fig. \ref{fig:histogram} presents the comparison of the proposed method with the standard AC-OPF, highlighting total cost and computational time. Notably, the proposed method achieves a 100\% feasibility ratio, meaning it consistently identifies operating points that are feasible within the standard AC-OPF framework. This outcome is achieved despite the challenges in accurately representing the non-convex feasible regions of the DSs through the NN classification model. The effective use of weights during the NN training process, along with the regression models that create a physical coupling between DS-related variables and PCCs, enables a holistic definition of the decision boundary, thereby preventing the proposed method from misclassifying infeasible points as feasible.

Examining Fig. \ref{fig:histogram} reveals that the average cost difference is 1.01\%, with only 40 out of 1,000 analyses showing a cost difference exceeding 2\%. This indicates that the proposed method, which prioritizes data privacy, achieves acceptable cost differences with minimal trade-offs. It is also noteworthy that the use of a high weight value during the NN training process to achieve a 100\% feasibility ratio results in a conservative approximation, causing a slightly higher cost difference. Additionally, another factor influencing the marginally higher cost difference is the use of a relatively small TS to test the proposed method under stricter conditions. This setup amplifies the impact of the three DSs on the total cost, making the relative cost difference appear more pronounced. Regarding computational efficiency, the average time difference is only 0.0639 seconds, with a maximum difference of 0.1235 seconds, underscoring the minimal time overhead introduced by the proposed method. This efficiency is attributed to the novel NN architecture used in the proposed method, which enhances computational speed.

Moreover, the proposed method effectively accommodates diverse FPU characteristics while maintaining strong performance. Given that different FPU characteristics are treated as constraints, this approach demonstrates the capacity to incorporate other potential market-based or operational constraints between TSOs and DSOs without any loss in performance. Consequently, the proposed method achieves a balance between data privacy and operational efficacy, yielding comparable performance to the standard AC-OPF. This capability enables DS flexibility to be leveraged for network management purposes without compromising data privacy.

\section{Conclusion}\label{sec:conclusion}

With the transformation of the power system, distribution systems (DSs) are playing an increasingly crucial role, accompanied by a growing number of flexibility-providing units (FPUs). Leveraging the flexibility offered by DSs has become essential for ensuring that network management is both cost-effective and secure. Achieving this requires seamless interoperability among network stakeholders, including Transmission System Operators (TSOs) and Distribution System Operators (DSOs). However, concerns regarding the disclosure of sensitive information, such as network topology and customer load profiles, hinder this interoperability and impede effective network management.

In this context, we propose a machine learning (ML)-based method in the present paper that prevents sensitive data from circulating between stakeholders, thereby enhancing interoperability across the network. In our approach, we represent the technical constraints of the DSs using ML models, which can be shared with the TSO without compromising data privacy. By leveraging these ML models, the TSO can solve the optimal power flow (OPF) problem and directly determine the dispatch of FPUs. This allows for dispatch decisions to be made in a single round of communication, eliminating the need for an additional disaggregation step. Furthermore, we demonstrate the method's flexibility by applying it to FPUs with a variety of PQ characteristics, not limited to ideal rectangular PQ charts, indicating that the method is adaptable to diverse FPU characteristics. Additionally, the flexibility potential of DSs is leveraged more effectively by accounting for variations at points of common coupling (PCCs) voltage. Moreover, to accurately represent the feasible region of the DSs, we propose a novel, tailored neural network (NN) architecture that performs this task with high computational efficiency.

The proposed method is benchmarked against the standard AC-OPF using multiple DSs with meshed connections and multiple PCCs. The results demonstrate high performance in terms of ML accuracy and overall effectiveness, highlighting the capability of the proposed method to protect data privacy while achieving reliable results. By modeling DSs with ML models, the TSO is prevented from accessing sensitive DS information, allowing the flexibility from DSs to be leveraged in network management without compromising data privacy. This approach thus promotes interoperability among stakeholders and enables more effective and secure network management.

\bibliographystyle{IEEEtran}
\bibliography{main.bib}

\vfill

\end{document}